\documentclass[conference,10pt,a4paper]{IEEEtran}
\IEEEoverridecommandlockouts
\usepackage{cite}
\usepackage{amsmath,amssymb,amsfonts}
\usepackage{algorithmic}
\usepackage{graphicx}
\usepackage{rotating}
\usepackage[leftcaption]{sidecap}
\usepackage{textcomp}
\usepackage[table,dvipsnames]{xcolor}
\usepackage{etoolbox}

\graphicspath{{figures/}}

\usepackage{listings}
\definecolor{codegray}{rgb}{0.5,0.5,0.5}

\definecolor{maroon}{rgb}{0.5,0,0}
\definecolor{darkgreen}{rgb}{0,0.5,0}

\lstdefinelanguage{XML}
{
  basicstyle=\ttfamily\tiny,
  morestring=[s]{"}{"},
  morecomment=[s]{?}{?},
  morecomment=[s]{!--}{--},
  commentstyle=\color{darkgreen},
  moredelim=[s][\color{black}]{>}{<},
  moredelim=[s][\color{red}]{\ }{=},
  stringstyle=\color{blue},
  identifierstyle=\color{maroon}
}

\ifCLASSOPTIONcompsoc
\usepackage[caption=false,font=normalsize,labelfont=sf,textfont=sf]{subfig}
\else
\usepackage[caption=false,font=footnotesize]{subfig}
\fi

\def\BibTeX{{\rm B\kern-.05em{\sc i\kern-.025em b}\kern-.08em
    T\kern-.1667em\lower.7ex\hbox{E}\kern-.125emX}}
    
\usepackage{algorithmic}

\usepackage[linesnumbered,ruled,noend]{algorithm2e}

\usepackage{tikz}

\usepackage{verbatim}

\usepackage{dirtytalk}

\begin{document}

\title{Full Feature Spiking Neural Network Simulation on Micro-Controllers for Neuromorphic Applications at the Edge}

\author{\IEEEauthorblockN{
Lars Niedermeier\IEEEauthorrefmark{1}, Jeffrey L. Krichmar\IEEEauthorrefmark{2}\IEEEauthorrefmark{3}}
\IEEEauthorblockA{\IEEEauthorrefmark{1}\textit{Niedermeier Consulting, Zurich, ZH, Switzerland}}
\IEEEauthorblockA{\IEEEauthorrefmark{2}\textit{Department of Computer Science}, \textit{University of California, Irvine, CA, USA}}
\IEEEauthorblockA{\IEEEauthorrefmark{3}\textit{Department of Cognitive Sciences}, \textit{University of California, Irvine, CA, USA}}
\IEEEauthorblockA{Correspondence Email: lars@niedermeier-consulting.ch}  
}

\maketitle

\begin{abstract}
Microcontroller units (MCU), which have an order of magnitude lower Size, Weight and Power (SWaP) than standard computers, makes them suitable for applications at the edge. Neuromorphic computing, which can realize low SWaP, relies on Spiking Neural Networks (SNNs). 
Until now, software based simulations
of SNNs required GPU-based workstations, application classified core processors such as the ARM Cortex-A53, or specialized hardware like Intel's Loihi. 
In the present work, we demonstrate that the SNN simulator CARLsim can run its full feature set on a MCU RP2350 with 8 MB memory.
We accomplished this by utilizing IEEE 16-bit float point numbers, which reduced memory requirements without loss of function. 
We were able to run the Synfire4 benchmark which comprises 1200 neurons. The accuracy was 97.5\% compared to the standard single precision numbers.  
Furthermore, we show that CARLsim runs a Synfire4 benchmark scaled-down to 186 neurons on a MCU in real-time at only 20 mW. 
Compared to the smallest application class ARM processor used by Raspberry in their Pi Zero 2 W, our MCU implementation is five times more energy efficient for the SNN itself, and an order of magnitude better when compared to the complete SoC (MCU/CPU + Board).

\end{abstract}

\begin{IEEEkeywords}
Edge Computing,
MCU, 
Neuromorphic Applications,
Spiking Neural Networks,
SWaP.
\end{IEEEkeywords}

\section{Introduction}
\label{sec:intoduction} 
 
Neuromorphic applications running at the edge require compute nodes with extreme low SWaP.  
This is true in particular for bioinspired applications that run Spiking Neural Networks (SNNs) for processing sensory input, 
controlling actuators by inference, and continuously learning and adapting.
SNNs require either special neuromorphic chips \cite{davies2018loihi, furber2014spinnaker, indiveri2011neuromorphic} or advanced SNN simulation software \cite{bekolay2014nengo, goodman2008brian} on power efficient general purpose hardware, such as mobile devices. 

CARLsim is an SNN simulation framework that supports computational neuroscience research and neuromorphic applications \cite{Niedermeier2022}. It is highly optimized for GPUs and multi-core CPUs. Its features include multiple neuron models, synaptic currents, learning rules, and visualization tools. With the introduction of a multi-threading kernel, CARLsim scales from energy efficient mobile SoC CPUs to datacenter GPUs \cite{niedermeier2026iscas}. 

However, MCUs are even more energy efficient (around an order of magnitude) than those CPUs, 
as they are used in ultra low energy demanding use-cases such as a TV remote control 
and numerous Internet-of-Things (IoT) devices.
The RP2350, the latest MCU developed by Raspberry, has an ARM Cortex-M33 which is more than 66\% faster   
than its predecessor, the RP2040, that is based on the ARM Cortex-M0, \cite{arm_cortex_m}. 
This motivated us to investigate if CARLsim might run on these MCUs, which would open up the potential for a new class of applications based on SNNs at the edge.   

\begin{figure}[ht]
 \centering
 \includegraphics[width=1.00\columnwidth]{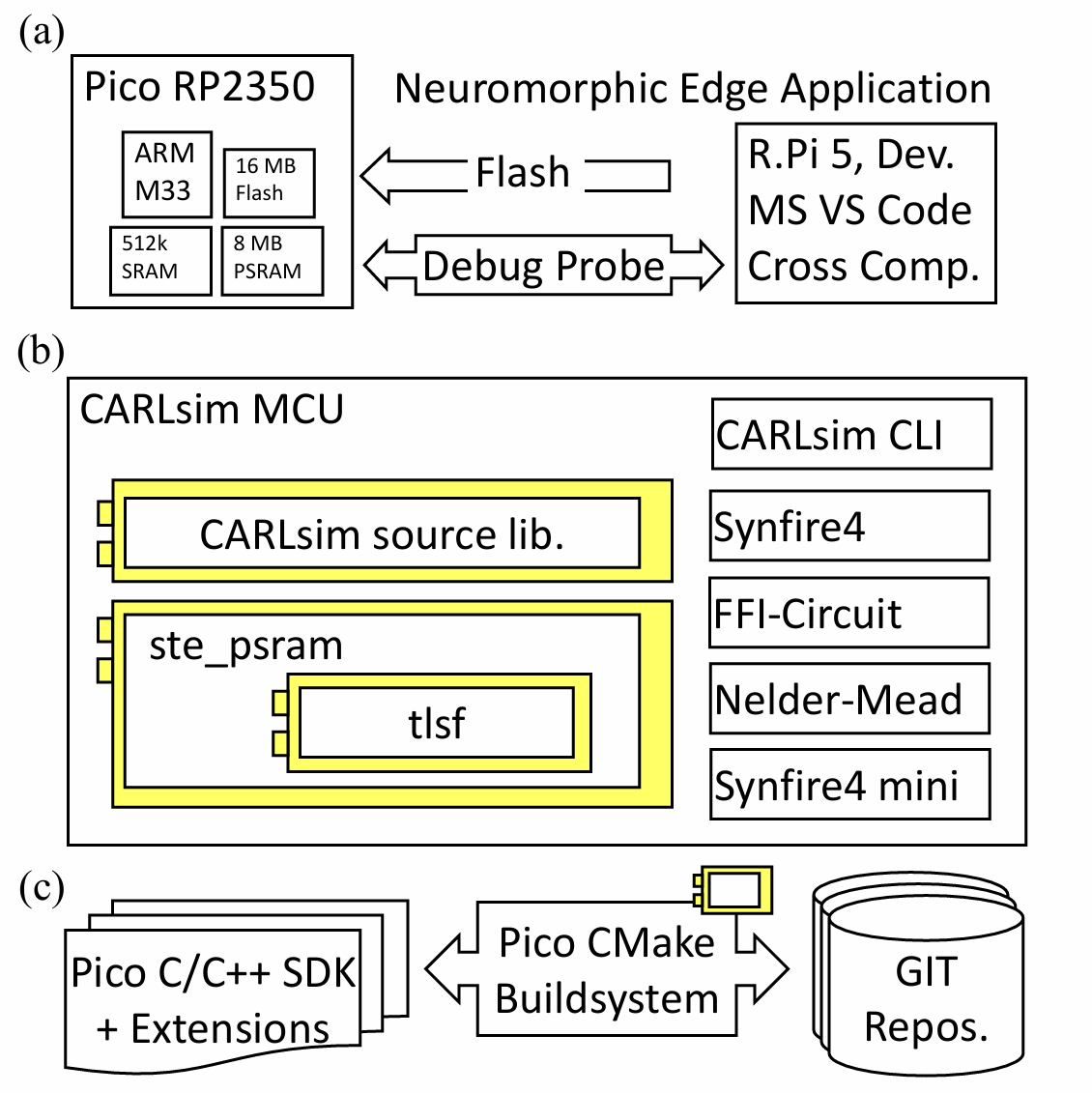}  
 \caption{Neuromorphic Applications on RP2350 MCUs. 
 (a) An MCU has no operating system (OS) with a hardware abstraction layer and one dedicated application running that is \emph{flashed} to its persistent memory. 
 (b) CARLsim MCU enables a new class of neuromorphic applications. 
 (c) The Raspberry Pico build system enables reusable source library based on CMake such as the SparkFun extension for the 8 MB PSRAM memory extension.  
}
\label{fig:intro_carlsim_mcu_system}
\end{figure}

While the Raspberry Pi Pico 2 gave us a ramp-up start with its well-established open-source ecosystem for the RP2350, 
it also quickly confirmed our concern that its 512 kB static random-access memory (SRAM) was insufficient to run CARLsim with its rich set of functionality and consequently large C++ code segment. 
However, the RP2350  does support a memory extension of up to 32 MB of pseudo SRAM (PSRAM).
It also turned out, that at the time of this writing, a few vendors such as Pimoroni and SparkFun offer such an implementation of Raspberry Pico's open hardware specification for the RP2350 with such an PSRAM extension. We concluded that SparkFun Pro Micro \cite{SparkFun_ProMicro_RP2350} with its 8 MB PSRAM 
was the most promising candidate to run CARLsim. 
The used PSRAM by APmemory is high speed, has low pin count interface which comprises of four SDR I/O pins, and
operates in quad peripheral interface (QPI) mode with frequencies up to 133 MHz \cite{APM_SPI_3V_PSRAM_Datasheet, apmemory_qspi_psram_datasheet}.

In the present work, we introduce a MCU implementation for CARLsim and demonstrate it on 
commercially available Raspberry Pico RP2350 boards (see Figure \ref{fig:intro_carlsim_mcu_system}). 
All code and models are open-source and available on GitHub \cite{CARLsim6Repository}.  
The main contributions of this work are:

\begin{enumerate}
\item \textbf{\textit{8MB PSRAM.}} Integration of the custom memory handler to utilize SparkFun's 8 MB PSRAM in CARLsim.
\item \textbf{\textit{FP16.}} Redesign the single float based storage of synapses as FP16 to address the limited RAM of MCUs. 
\item \textbf{\textit{Pico SDK extension.}} Implement CARLsim as source library extension for the Pico C/C++ SDK.
\item \textbf{\textit{Reference implementation.}} Customize the RP2350 Pro to support the Raspberry debug probe in VS Code.
\item \textbf{\textit{Performance benchmark.}} Apply the Synfire4 benchmark to measure the actual SNN performance.
\item \textbf{\textit{Sizing for real-time.}} Downsize Synfire4 for real-time use cases and demonstrate power efficiency of MCUs over CPUs.  
\end{enumerate}

\section{Methods}
\label{sec:methods} 

\subsection{Spiking neuron model}
\label{subsec:methods_neuron_model}

CARLsim efficiently implements spiking neuron models such as the leaky-integrated-and-fire (LIF) and the Izhikevich neuron model with 4 and 9 parameters (IZH4, IZH9) \cite{Niedermeier2022}. In the present work, we use the biologically plausible IZH4 to implement the Synfire network \cite{niedermeier2026iscas}. It is described by the following equations \cite{Izhikevich2003}.
\begin{eqnarray} \label{eq:Izhi4}
\dot{v} &{}={}& 0.04v^2 + 5v + 140 - u + I    \\
\dot{u} &{}={}& a(bv - u)  \label{eq:Izhi4u}
\end{eqnarray}
\begin{equation} \label{eq:Izhi4v30}
\text{if}\ v \geq 30 \begin{cases}
v = c  \\
u = u + d
\end{cases}
\end{equation}

\subsection{Raspberry Pico RP2350 MCU}

\begin{figure}[htb]
\centering
\hfill
\subfloat[RP2350 MCU]{\includegraphics[width=0.425\columnwidth]{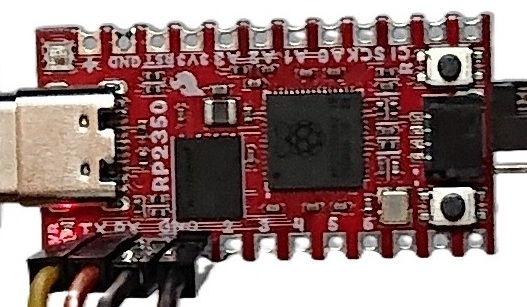}\label{subfig:method_rp2350_mcu}} 
\hfill
\hfill
\subfloat[Backside]{\includegraphics[width=0.425\columnwidth]{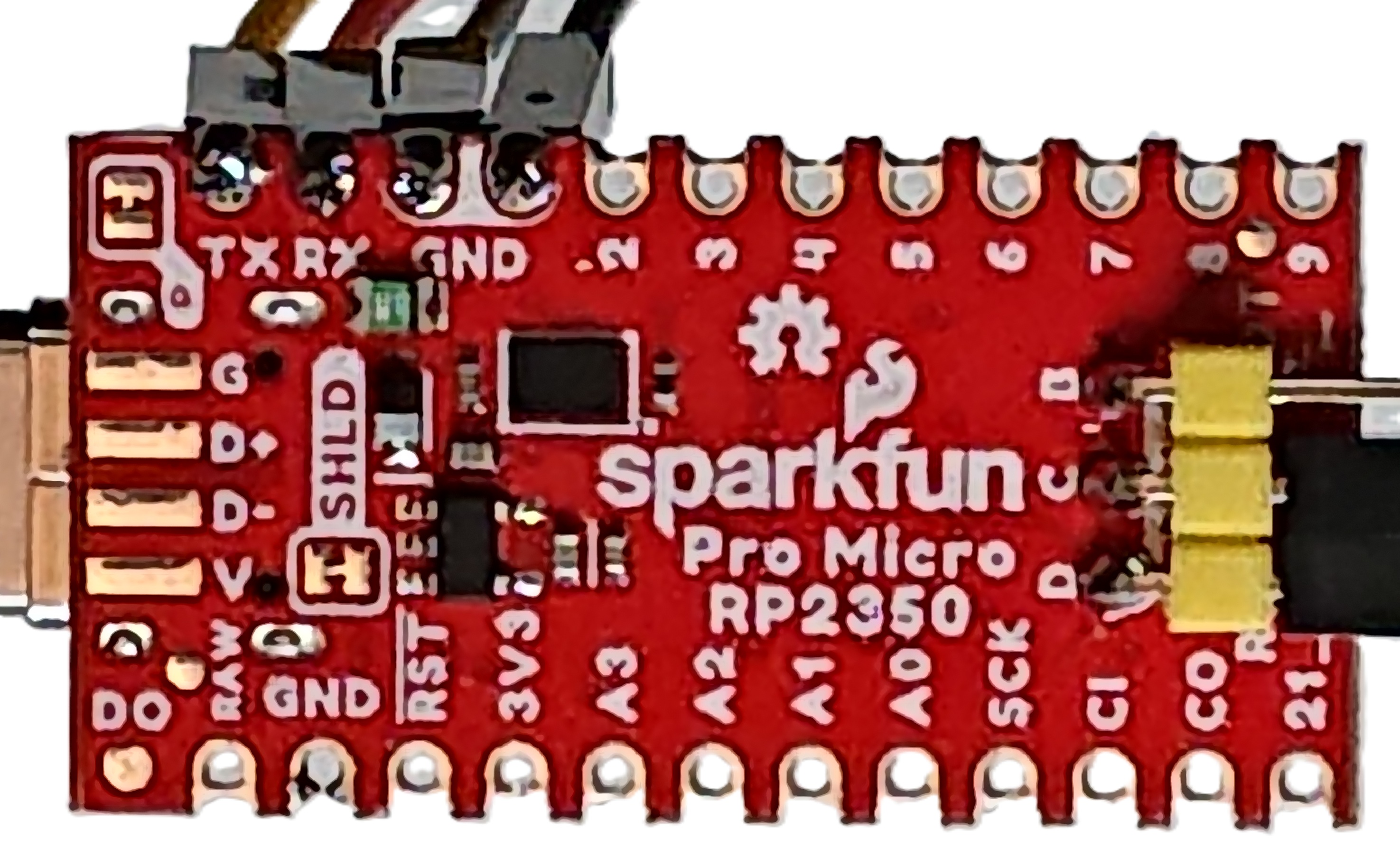}\label{subfig:method_rp2350_swd_connector}}
\hfill
\hfill
\caption{
The RP2350 MCU with two ARM Cortex-M33 cores. 
a) The SparkFun Pro Micro RP2350 has 8MB PSRAM extension (on the backside) that enables CARLsim.
b) RP2350 MCU (backside) with 8MB PSRAM. The chip APS6404L, package code USON-8L, by apmemory is only 3x2mm small (left above \emph{sparkfun}).  
   The SWD connector for the debug probe was a custom extension soldered onto the test points of the board.}
\label{fig:method_rp2350_mcu}
\end{figure}

We selected the Raspberry Pico RP2350 MCU for our reference implementation of CARLsim 
due to its dual ARM Cortex-M33 and its open source and hardware policy.
Its processors also can be switched to the open instruction set Hazard3 RISC-V. 
Multiple vendors offer customized products compatible 
with the reference implementation in the range of \$5 to \$20. 
As we anticipated, the 512 kB static random access memory (SRAM) of the reference implementation 
was insufficient for CARLsim. 
Therefore we selected the SparkFun RP2350 Pro Micro as it integrates 8 MB pseudo static random access memory (PSRAM)
and officially supports the Pico C/C++ SDK development environment.
Figure \ref{fig:method_rp2350_mcu} shows the SparkFun Pro Micro RP2350. 
The USB-C connector at the left side gives a visual reference of its small dimensions.
It has the smallest SWaP (33.02mm x 17.78mm) which is about half of the size of a Pimoroni Pico LiPo 2 (53mm x 21mm).   

\subsection{Raspberry Pico C/C++ SDK} 

Since MCUs do not have an operating system or a development tool chain on the board, MCUs have extremely low SWaP.  Applications are cross-compiled on a host system such as an Raspberry Pi 5. 
The MCU is then \emph{flashed} with the application which will be running exclusively on the MCU (see Figure \ref{fig:intro_carlsim_mcu_system}a).

Raspberry recommends Microsoft Visual Studio (VS) Code as development IDE for the Pico 2  
and provides \emph{Raspberry Pi Pico 0.19.0} as the official VS Code extension for the Raspberry Pi development. 
It enables the developer to conveniently use the Pico C/C++ SDK, for  project generation and setup, 
installation of dependencies, automatic CMake configuration, version switching, one-click compilation and debugging. 
For the full feature set refer to the documentation \cite{raspberry_pico_sdk, raspberry_pico_getting_started, raspberry_pico-vscode_repo}.

CMake was chosen as the defacto-standard for build-system both by Raspberry and VS Code. Conveniently, CMake was introduced for SNN development in CARLsim 6 \cite{Niedermeier2022}. 
As mentioned above, MCUs have a very different way to build and deploy software. 
For this reason, reusing static or dynamic build libraries for multiple applications does not  make sense. 
Hence, Raspberry utilizes CMake based source libraries that are similar reusable. 
CARLsim CMakeLists.txt are modified accordingly.

In our experience, the Raspberry Pi 5 worked well as a host development system.
Also, remote debugging of the PSRAM memory extensions and FP16 (see later sections) 
over serial wire worked reliably and fast.   
Therefore, we refrained from the Windows based VS Code as cross-compiling environment, 
as the debugger showed spurious behavior. 

\subsection{8 MB PSRAM}

\begin{figure}[ht]
 \centering
 \includegraphics[width=2.2in]{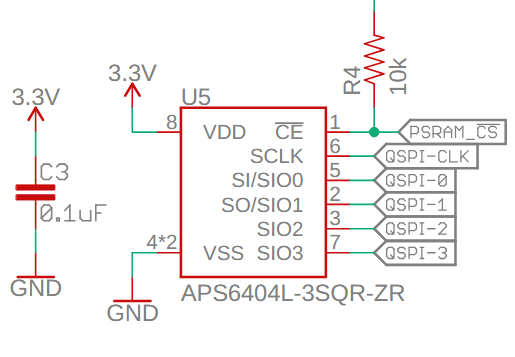}  
 \caption{PSRAM circuit.  Schematic of 8 MB PSRAM circuit for the APS6404L-3SQR-ZR QSPI chip.}
\label{fig:method_rp2350_psram_circuit_schematic}
\end{figure}

The key for enabling CARLsim on MCUs is the memory extension from 512 kB to 8.5 MB. 
Figure \ref{fig:method_rp2350_psram_circuit_schematic} shows the open source hardware schematic of the PSRAM chip \cite{SparkFun_ProMicro_RP2350_Schematic}. 
The APS6404L is a PSRAM device that features a high speed, low pin count interface and organizes 64Mb in 8M x 8bits \cite{apmemory_qspi_psram_datasheet}. 
It has four SDR I/O pins that are connected to the RP2350's Quad-I/O SPI (QSPI) controller and operate at a frequency up to 133 MHz.  
It incorporates a seamless self-managed refresh mechanism and does not require the support of DRAM refresh from a system host. 
Interestingly, exploring QSPI specification reveals that 64 Mb is currently the highest density for 3.0V available on a single chip \cite{apmemory_qspi_psram}. 

The \emph{sparkfun\_pico} library replaces the default allocator of the Pico SDK with a version that accesses both the build heap SRAM and PSRAM \cite{sparkfun_pico}.
It leverages the wrap functionality provided by the Pico SDK and hooks into the malloc/free functions of the C-library and the new/delete operators of the C++ standard library.
To enable this functionality, the following entries in the CMakeLists.txt file are necessary:

\begin{lstlisting}[language=Make]
# Enable the system malloc/free to be wrapped.
set(SKIP_PICO_MALLOC 1)

# Wrap sparkfun_pico builds.
set(SFE_PICO_ALLOC_WRAP 1)

# Wrap system malloc/free during compilation.
add_definitions(-DSFE_PICO_ALLOC_WRAP)

# Add the sparkfun_pico library.
add_subdirectory(sparkfun_pico)

\end{lstlisting}

SparkFun embeds \emph{tlsf}, a two-level segregated fit (TLSF) memory allocator implementation \cite{tlsf_spec}.
It features O(1) cost for malloc, free, realloc, memalign, extremely low overhead per allocation (4 bytes), 
low overhead per TLSF management of pools, low fragmentation, and compiles to only a few kB of code and data \cite{tlsf_repo}.

\begin{lstlisting}[language=C++]
// memory statistics based on SparkFun Pico sfe_ library
void mcuMemGetInfo(size_t *size, size_t *used, size_t *avail)
{
    *size = sfe_mem_size();
    *used = sfe_mem_used();
	*avail = sfe_mem_max_free_size();
}

// example usage in CARLsim 
void SNN::allocateSNN_CPU(int netId) {
    ...
    size_t size, avail, total, previous;
    float toMB = std::pow(1024.0f, 2);
    mcuMemGetInfo(&size, &total, &avail);
    KERNEL_INFO("CPU Memory Management: (Total %2.3f MB)",(float)(size/toMB));
    KERNEL_INFO("Data\t\t\tSize\t\tTotal Used\tTotal Available");
    ...    
\end{lstlisting}

\subsection{Debug Probe}

SparkFun strictly designed the Pro Micro RP 2350 for low SWaP. Its Quiic connector \ref{fig:method_rp2350_mcu} is an important selling proposition 
which allows daisy chain connect a whole ecosystem of compliant components such as sensors, actuators, and displays via the Inter-Inter-Chip (I2C) protocol. 

Due to the space limitation, the vendor prioritized the Quiic connector and sacrificed the connector for serial wire debug (SWD) \cite{serial_wire_debug} 
for the Raspberry Pi debug probe \cite{debug_probe}.
Thanks to the open hardware policy \cite{SparkFun_ProMicro_RP2350},
the low-level debug protocols \cite{arm_debug_spec, cmsis_dsp} are supported via the SWD interface, using test points TP1 for SWDCK and TP2 for SWDIO.
The schematic \cite{SparkFun_ProMicro_RP2350_Schematic} shows that the Quiic connector links GPIO pins 16 and 17 (SDA and SCL), instead to SWDCK (TP1) and SWDIO (TP2) which are the outer contacts of an SWD plug.
The test points TP1 and TP2 are exposed as D and C at the bottom of the printed circuit board (PCB).
The connector was soldered onto the test points and secured to the PCB with adhesive to withstand physical stress.
 
\subsection{FP16}

With the setup shown in Figure \ref{subfig:method_rp2350_swd_connector}, 
we had our development environment working, 
and promptly got a heap fault as we were still 2 MB short on memory.
We then optimized several aspects of CARLsim, but most important of all, 
we implemented FP16 floating point arithmetic, rather than 32 bit floating points, to reduce memory storage. With FP16, CARLsim can load and run the Synfire4 benchmark on the RP 2350 MCU. 
FP16 uses ten bits for the fraction (mantissa) and five bits for the exponent, 
in contrast to IEEE 754 single float, which have 23 bits for the fraction and eight bits for the exponent, \cite{exxactcorp2026fp32fp16}
\cite{wikipedia2026fp16}. 

\begin{lstlisting}[language=C++] 
// synaptic storage at RUNTIME state 
typedef struct RuntimeData_s {
    ...
    __fp16* wt;       //!< stores the weight change of a synaptic connection
    __fp16* wtChange; //!< stores the weight change of a synaptic connection
    __fp16* maxSynWt; //!< maximum synaptic weight for a connection
    ...

// synaptic storage at SETUP state 
typedef struct ConnectionInfo_s {
    short int grpSrc;      	
    short int grpDest;     
    uint8_t delay;	        
    __fp16 initWt;
    __fp16 maxWt;
    ...
\end{lstlisting}

To be able to build with FP16 support, the following compiler flags needs to be defined for CMake. 
\begin{lstlisting}[language=Make]
CMAKE_CXX_FLAGS:STRING=-mfp16-format=ieee -mcpu=cortex-m33 -mthumb \\
                 -march=armv8-m.main+fp+dsp -mfloat-abi=softfp -mcmse
\end{lstlisting}

\subsection{Synfire4 benchmark}
\label{subsec:methods_synfire4_benchmark} 

\begin{figure}[ht]
\centering
\includegraphics[width=0.95\columnwidth]{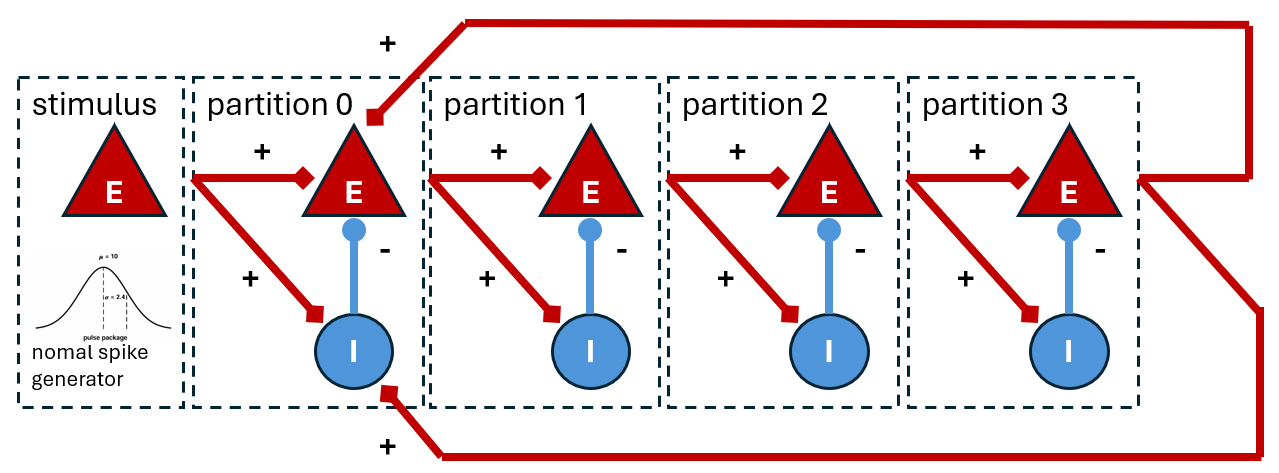}
\label{fig:method_synfire_carlsim}    
\caption{Architecture of Synfire4 benchmark network. The four segments are recurrently connected to measure the performance of a typical traveling wave through the SNN. The normal spike generator can generate various types of stimulus pulses.
}
\label{fig:method_synfire4}
\end{figure}

We introduced the Synfire chain network as a benchmark to measure and compare performance
of CARLsim and neuromorphic chips in \cite{niedermeier2026iscas}. We followed Höppner et. al. in their approach for SpiNNaker \cite{hoeppner2022} and built a network with the same structure and sizing (see Fig. \ref{fig:method_synfire4}). An  investigation of Synfire benchmark revealed that LIFs with their arbitrary threshold as activation function produce poor results for correlated inhibition \cite{niedermeier2026iop}. Therefore, instead of LIFs we used the IZH4 neuron model in our SNN. 
Figure \ref{fig:method_synfire4} presents the Synfire benchmark with 4 segments (Synfire4) 
we used for measuring the energy efficiency, see section \ref{sec:Results}. Results. 

Table \ref{table:Synfire4Neurongroups} presents the parameters for the neuron groups. Each excitatory group $E$ has 200 regular spiking (RS) neurons $(a=0.02, b=0.2, c=-65, d=8)$, and the inhibitory groups consist of 50 fast spiking (FS) neurons $(a=0.1, b=0.2, c=-65, d=2)$. The connection groups are defined in Table \ref{table:ConnectionGroups}. The code and configuration is open source and available in the CARLsim GitHub repository \cite{CARLsim6Repository}. 

\begin{table}[ht]
\caption{Izhikevich neuron parameter for Synfire SNN with four segments.}
\label{table:Synfire4Neurongroups}
\centering
\begin{tabular}{lrlrrrr}
\hline
\bf Group& \bf Size& \bf Neuron Type&	 \bf a& \bf b& \bf c& \bf d\\ 
\hline
Cstim& 	200&    poisson (spike generator)&   &	& 	&	\\ 
Cexc[i] & 	200&	excitatory (regular spiking)&   0.02&	0.2& 	-65&	8\\ 
Cinh[i] & 	50&    inhibitory (fast spiking)& 	  0.1&	0.2&	-65&	2\\
\hline
& & $i \in \{0,1,2,3\}$
\end{tabular}
\end{table}

\begin{table}[ht]
\caption{Synaptic connections of Synfire SNN with recurrent four segments.}
\label{table:ConnectionGroups}
\centering
\begin{tabular}{llrrr}
\hline
\bf Pre-Synaptic & \bf Post-Synaptic& \bf Connections*& \bf Weight& \bf Delay\\   
\hline

Cstim& 	Cexc[0]& 60&   1.0&	10 ms\\ 
Cstim& 	Cinh[0]& 60&   3.5&	10 ms\\ 

Cexc[i]& 	Cexc[i+1]& 60&   1.0&	10 ms\\ 
Cexc[i]& 	Cinh[i+1]& 60&   3.5&	10 ms\\ 
Cinh[i+1]& 	Cexc[i+1]& 25&   -2.0&	8 ms\\ 

Cexc[3]& 	Cexc[0]& 60&   1.0&	10 ms\\ 
Cexc[3]& 	Cinh[0]& 60&   3.5&	10 ms\\ 

\hline
& $i \in \{0,1,2\}$ & * per neuron 
\end{tabular}
\end{table}

\section{Results}
\label{sec:Results}

\subsection{Performance on the Synfire4 benchmark}

We tested the performance of CARLsim on a MCU SparkFun RP2350 Pro Micro with the same Synfire4 benchmark 
we have applied for the CARLsim OMP kernel on Raspberry Cortext-A processors such as the Raspberry Pi Zero 2 W \cite{niedermeier2026iscas}.
As described in the Methods section \ref{subsec:methods_neuron_model}, 
the Synfire4 benchmark is a feed-forward-inhibition SNN with 1.2k neurons and roughly 81k synapses.
We applied this size of SNN as it represents a reasonable upper memory limit for MCUs. 
The Cortex-M33 processing core of an RP2350 MCU is much slower than the Cortex-A53 which is used in the smallest Raspberry Pi. 
To measure the performance of CARLsim on MCUs, we measured the time that the simulation took in relation to a wall clock.
In addition, we measured the power consumption with a USB meter. 

The main goal was to demonstrate that a full-scale SNN simulator like CARLsim can run on an MCU. This would show that low SWaP neuromorphic applications could be realized on  a completely different class of compute hardware than CPUs, GPUs and specialized hardware.  
Unlike CPUs, MCUs lack a hardware abstraction layer (HAL), 
an operating system (OS), and supplemental hardware for I/O such as keyboard, video, and mouse (KVM). 
Table \ref{table:Synfire4Memory} shows the memory ramp-up for the Synfire4 benchmark on the RP2350. 
The total available SRAM and PSRAM memory is 8.5 MB. 
There are seven steps for CARLsim to load and run an SNN: Initialize the internal data structures, 
prepare the random generators and connection info, optimize the runtime memory to hold the synapses and neurons states, 
and finalize the auxiliary data such as spike tables. 
The values presented in the table are copied from the terminal output that CARLsim produces in accordance with the set log level. 
The first column shows the memory size of the particular step, the second the incremental memory use, and the third 
the decrease in available memory. CARLsim internally calls the functions of the hooked memory allocator  
that reuses, caches, and reorganizes the heap according to the heuristics defined by TLSF \cite{tlsf_spec, tlsf_repo}. 
This gives a practical indication of the effective memory usage. 
As a consequence, the numbers cannot just be summed and simply compared. 

\begin{table}[ht]
\caption{ Memory ramp-up of CARLsim MCU for the Synfire4 benchmark (1200 neurons) on a SparkFun Micro Pro 2 RP2350.}   
\label{table:Synfire4Memory}
\centering
\begin{tabular}{lrrr}
\hline
\bf Simulation load step& \bf Mem. Size& \bf Total Used&	 \bf Total Available\\ 
\hline
(SRAM + PSRAM) &                   &                &        8.477 MB\\                                    
1. CARLsim Init.&          4.228 MB&        4.228 MB&        4.249 MB\\  
2. Random Gen.&            0.001 MB&        4.229 MB&        4.248 MB\\
3. Conn. Info&             1.315 MB&        5.544 MB&        2.934 MB\\
4. Syn. State&             0.156 MB&        5.700 MB&        2.777 MB\\
5. Neuron State&           0.082 MB&        5.782 MB&        2.696 MB\\
6. Group State&            0.141 MB&        5.922 MB&        2.555 MB\\
7. Auxiliary Data&         1.665 MB&        7.587 MB&        0.890 MB\\
\hline
\end{tabular}
\end{table}

The simulation was executed for 1 s model time and produced 27,364 spikes (vs. 26,694 spikes of a reference run using single-precision floats).
The average firing rate was 22.8 Hz. Running the simulation for 1 second model time took 27 seconds wall clock time. 
As the metric for the correctness of the SNN simulation, we compared the number of spikes on the RP2350 
to a reference simulation on the CPU based Raspberry Pi Zero 2 W. 
The simulation ran at an accuracy of 97.5\% compared to the single precision floats.

\subsection{Energy efficiency for real-time use-cases}

\begin{figure}[ht]
 \centering
 \includegraphics[width=0.9\columnwidth]{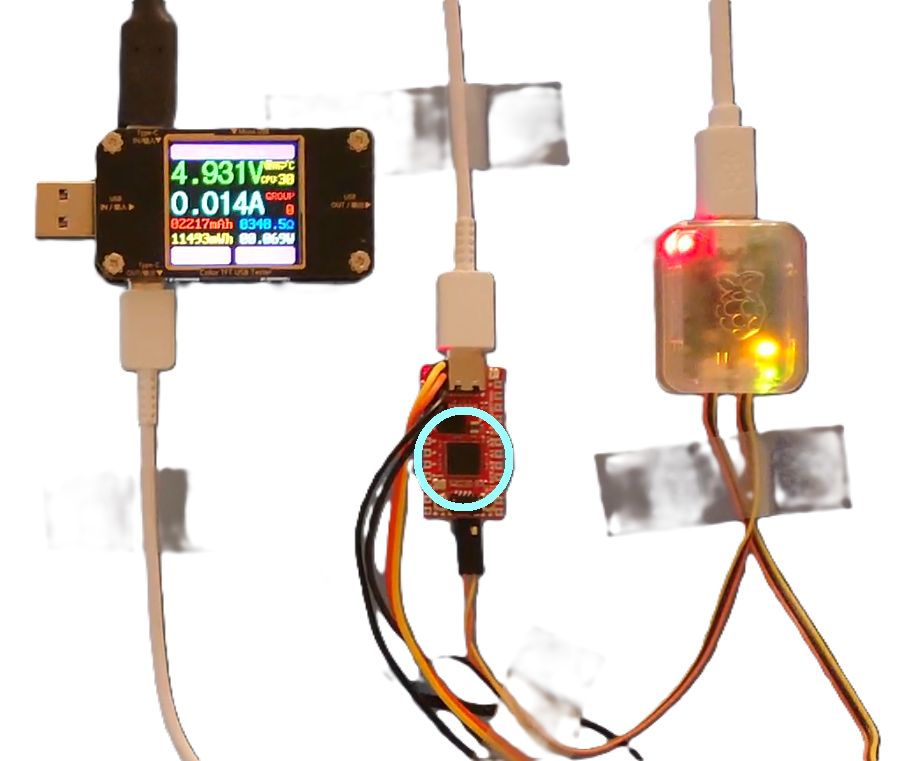}
 \caption{CARLsim Synfire4 benchmark scaled-down to 186 neurons for running Synfire SNN in real-time on a RP2350 MCU (marked by the cyan circle) requires less than 20 mW of power. The baseline of the MCU is about 50 mW.
 The power was measured with the USB power meter at the left side. 
 At the right side, the connected Raspberry Pi Debug Probe is shown, over which the experiment was controlled.}
\label{fig:result_rp2350_mcu_synfire4_196_rt}
\end{figure}

Figure \ref{fig:result_rp2350_mcu_synfire4_196_rt} shows the setup for running CARLsim on the Synfire4 benchmark scaled-down to 186 neurons running in real-time.  
Table \ref{table:Synfire4miniMemory} shows the memory ramp-up of CARLsim MCU for the Synfire4 benchmark mini (186 neurons) on a SparkFun Micro Pro 2 RP2350. 
The numbers clearly demonstrate that significantly less memory is used for the mini SNN. This means the bottleneck for the mini SNN is not the memory 
but, with respect to real-time processing, the processing power. 
Utilizing the second core of the RP2350 could therefore support a Synfire4 SNN twice the size (372 neurons). We estimated 2.4 MB of memory usage which leaving plenty of reserve. 
In the  Conclusion section, we will briefly discuss further measures that have the potential to enable a Synfire4 SNN of 1k neurons to run in real-time. 
We estimate a memory reserve of approximately 1.2 MB will be left available for other usage.  
These considerations indicate that the proposed RP2350 MCU with the PSRAM extension is well-balanced for the use with CARLsim.  

Table \ref{table:Synfire4Performance} summarizes the results of both benchmarks. 
We measured the simulation time, effective current, voltage, and power utilized by the simulations with an off-the-shelf retail USB-Tester. 
The applied USB-Tester had an voltage (current) resolution of 0.01V (0.001A) and an accuracy of ±0.2\% (±0.8\%).
The baseline power usage of MCU was around 50 mW (0.010 A). 
Running the SNN requires less than 20 mW (0.014 A) of additional power. 

A metric for the energy efficiency often used in neuromorphic hardware studies is Joule per spike \cite{furber2014spinnaker}.
In the Synfire4 benchmark we have 27,364 spikes that use 30 mW over 27.4s so we have 30$\mu$J/spike for the SNN simulation.  
The complete system uses 79$\mu$J/spike.  
At the time of writing, no suitable performance comparison to Loihi or SpiNNaker is available. 
SpiNNaker uses around 8nJ/spike for both Izhikevich and LIFs models but for networks having 192,000 neurons \cite{stromatias2013power}. 
Loihi has a dynamic energy efficiency of 23.6pJ/spike for several SNNs \cite{davies2018loihi} but also has around 1W system overhead \cite{blouw2019benchmarking}. 
A valid comparison requires experimental constrains such as the network size, the applied neuron and synaptic model,
the precision and kind of of the integration method (Euler or Runge/Kutta), 
and of course the frequency and the input strength.   

Videos in the supplemental materials  
show the complete runs of both benchmarks from start to finish including the setup over the SWD debug probe \cite{niedermeier2026wcci_supplemental_synfire4, niedermeier2026wcci_supplemental_synfire4mini}.  
The reference run on a Raspberry Zero 2 W with DCA on one A53 core confirms the expected improvement in energy efficiency   
of an order of magnitude for the overall current of the system (RP2350 MCU, Raspberry Zero 2 W PCB) and
of a factor of five for the power of the SNN\cite{niedermeier2026wcci_supplemental_synfire4mini_a53}. 

\begin{table}[ht]
\caption{Memory ramp-up of CARLsim MCU for the Synfire4 benchmark mini (186 neurons) on a SparkFun Micro Pro 2 RP2350.}   
\label{table:Synfire4miniMemory}
\centering
\begin{tabular}{lrrr}
\hline
\bf Simulation load step& \bf Mem. Size& \bf Total Used&	 \bf Total Available\\ 
\hline
(SRAM + PSRAM)&                   &                &        8.478 MB\\                                    
1. CARLsim Init.&          0.808 MB&        0.808 MB&        7.670 MB\\  
2. Random Gen.&            0.000 MB&        0.808 MB&        7.670 MB\\
3. Conn. Info&             0.047 MB&        0.855 MB&        7.623 MB\\
4. Syn. State&             0.005 MB&        0.860 MB&        7.618 MB\\
5. Neuron State&           0.012 MB&        0.871 MB&        7.606 MB\\
6. Group State&            0.108 MB&        0.979 MB&        7.499 MB\\
7. Auxiliary Data&         0.205 MB&        1.183 MB&        7.294 MB\\
\hline
\end{tabular}
\end{table}

\begin{table}[ht]
\caption{Performance comparison of CARLsim on the Synfire4 benchmark for SparkFun Pro Micro RP2350 (Arm Cortex-M33) vs. Raspberry Pi Zero 2 W (Arm Cortex-A53).}
\label{table:Synfire4Performance}
\centering
\begin{tabular}{lrrr}
\hline
\bf Description & \bf Synfire4&   \bf Synfire4,mini&  \bf Synfire4,mini\\ 
\bf  & \bf MCU (M33)&   \bf MCU (M33)&  \bf CPU (A53)\\ 
\bf  & \bf RP2350&   \bf RP2350&  \bf Pi Zero 2 W\\ 
\hline
SNN, neurons:&                1,200 &  186  &   186\\                                    
SNN, synapses:&              81,154 &  2,430  &  2,430 \\    
Execution time: &            27.4 s & 29.7 s   & 30.02 s\\
Model time:&       1 s & 30 s  & 30 s\\
Spike processed:&   27,364 & 412  & 412\\
Average Firing rate:  & 22.8 Hz &  0.074 Hz  & 0.074 Hz\\
Current (MCU/ SOC): &  0.016 A & \bf 0.014 A  & 0.148 A \\  
Power for SNN: &  30 mW  & \bf 20 mW  &  101 mW \\ 
\hline
\end{tabular}
\end{table}

\section{Conclusion} 
\label{sec:discussion}

The main goal of this work was to get the full-scale CARLsim software framework to run on a MCU, as this would open up the possibility of low SWaP neuromorphic edge applications on inexpensive, commercially available hardware.
Compared to CPUs and GPUs, MCU's intrinsic design and properties target minimal SWaP. 
The CARLsim implementation of a MCU uses only 20 mW of power. This corresponds to one or two standard LEDs \cite{vishay_universal_led_tlur540}. All CARLsim features, such as the Izhikevich neuron model with 9 parameters, conduction-based synapses, compartments and Runge-Kutta for high precise integration, STP, STDP, and neuromodulation can now run on MCUs.  
This could enable a new class of neuromorphic applications at the edge.  In this first implementation, SNNs  with a few hundred neurons and several thousand synapses run at real-time.  This is roughly the size of some species of insect brains.

We demonstrated that CARLsim on an MCU can load and run the Synfire4 benchmark with 1.2k neurons and 81k synapses. For SNNs, the  memory required for storing the synaptic connections is the limiting hardware factor and required implementation of FP16 for parameters and mathematical operations.  Future work will engage the second M33 core to support several hundreds neurons at run-time. We have also identified further optimization potential in the ARM Cortex-M instruction set. This could enable SNNs sized of 1k neurons at real-time in the near future.   

\bibliographystyle{IEEEtran}
\bibliography{references}

\end{document}